\documentclass[12pt,preprint]{aastex}
\begin{document}
\title{The Extreme Ultraviolet Deficit - Jet Connection in the Quasar 1442+101}
\author{Brian Punsly\altaffilmark{1}, Paola Marziani\altaffilmark{2}, Preeti Kharb \altaffilmark{3}, Christopher P. O'Dea\altaffilmark{4,5}
and Marianne Vestergaard \altaffilmark{6}}\altaffiltext{1}{1415
Granvia Altamira, Palos Verdes Estates CA, USA 90274 and ICRANet,
Piazza della Repubblica 10 Pescara 65100, Italy,
brian.punsly1@verizon.net} \altaffiltext{2}{INAF, Osservatorio
Astronomico di Padova, Italia}\altaffiltext{3}{Indian Institute of
Astrophysics, II Block, Koramangala, Bangalore 560034, India}
\altaffiltext{4}{Department of Physics and Astronomy, University of
Manitoba, Winnipeg, MB R3T 2N2 Canada}\altaffiltext{5}{School of
Physics \& Astronomy, Rochester Institute of Technology, Rochester,
NY 14623, USA} \altaffiltext{7}{Dark Cosmology Centre, Niels Bohr
Institute, University of Copenhagen, Juliane Maries Vej 30, DK-2100
Copenhagen $\mathrm{\phi}$, Denmark}

\begin{abstract}
In previous studies, it has been shown that the long term time
average jet power, $\overline{Q}$, is correlated with the spectral
index in the extreme ultraviolet (EUV), $\alpha_{EUV}$ (defined by
$F_{\nu} \sim \nu^{-\alpha_{EUV}}$ computed between 700\AA\, and
1100\AA\,). Larger $\overline{Q}$ tends to decrease the EUV
emission. This is a curious relationship because it connects a long
term average over $\sim 10^{6}$ years with an instantaneous
measurement of the EUV. The EUV appears to be emitted adjacent to
the central supermassive black hole and the most straightforward
explanation of the correlation is that the EUV emitting region
interacts in real time with the jet launching mechanism.
Alternatively stated, the $\overline{Q}$ - $\alpha_{EUV}$
correlation is a manifestation of a contemporaneous (real time) jet
power, $Q(t)$, correlation with $\alpha_{EUV}$. In order to explore
this possibility, this paper considers the time variability of the
strong radio jet of the quasar 1442+101 that is not aberrated by
strong Doppler enhancement. This high redshift (z = 3.55) quasar is
uniquely suited for this endeavor as the EUV is redshifted into the
optical observing window allowing for convenient monitoring. More
importantly, it is bright enough to be seen through the Lyman forest
and its radio flux is strong enough that it has been monitored
frequently. Quasi-simultaneous monitoring (five epochs spanning
$\sim 40$ years) show that increases in $Q(t)$ correspond to
decreases in the EUV as expected.

\end{abstract}


\section{Introduction}Powerful relativistic radio jets have been
associated with the most intrinsically luminous ultraviolet emitters
in the Universe, namely quasars, since their discovery over 50 years
ago. Only a small fraction of all quasars ($\sim$ 10\%) have strong
jets (radio loud quasars RLQs), and even fewer have radio lobes on
super-galactic scales, $\sim 1.7\%$ of all quasars have such
extended structure \citep{dev06}. However, until recently there has
been no observations of the jet launching region, so there has been
much speculation as to the mechanism that creates these jets. The
literature is filled with numerical models and theories
\citep{lov76,bla77,bla82,pun08}. There is no discriminant for
authenticity due to a lack of direct measurement of the jet
launching region. However, this has changed recently with the
examination of extreme ultraviolet (EUV) spectra shortward of the
peak of the spectral energy distribution (SED) at 1100\AA\,. The
quasar luminosity is widely believed to arise from the viscous
dissipation of turbulence driven by the differential rotational
shearing of accreting gas \citep{lyn71,sha73}. In numerical and
theoretical models, the highest frequency optically thick thermal
emission arises from the innermost region of the accretion flow and
its frequency is shortward of the peak of the SED \citep{zhu12}. The
EUV spectrum beyond the peak of the SED of quasars is the putative
Wien tail of the emission of the innermost thermal component of the
accretion flow that is adjacent to the central black hole
\citep{mar97,pun14}. If $\overline{Q}$ is the long term ($\sim
10^{6}$ years) time averaged jet power (as determined from radio
lobe emission and morphology), and $L_{bol}$ is the bolometric
thermal emission associated with the accretion flow, it was shown in
\citet{pun14,pun15} that jet efficiency, $\overline{Q}/L_{bol}$,
(which depends on a long term average) was correlated with the
deficit of EUV emission in RLQs relative to radio quiet quasars
quantified by $\alpha_{EUV}$ (the flux density scales as
$F_{\nu}\propto\nu^{-\alpha_{EUV}}$ and is computed between
1100\AA\, and 700\AA). This is the fundamental correlation and a
partial correlation analysis shows that the correlation that exists
between $\overline{Q}$ and $\alpha_{EUV}$ is spurious \citep{pun15}.
This provides the first connection between jet power and an
observable from a region that is likely contiguous or coincident
with the jet launching region in quasars.
\par This is a curious
circumstance since the plasma in the radio lobes that is used to
determine $\overline{Q}$ was ejected from the central engine $\sim
10^{6}$ years before the EUV emitting gas reached the environs of
the central black hole. Two parameters, $L_{bol}$ and $\alpha_{EUV}$
are ``real time" diagnostics of the quasar at the time of
observation and $\overline{Q}$ is a long term time average. Why
should these parameters be connected in what is likely a time
variable system? It was concluded based on a statistical argument
that the most logical explanation is that there is an underlying
real time connection between the instantaneous jet power, $Q(t)$,
and $\alpha_{EUV}$ \citep{mar15}. It is too much of a coincidence
that two highly dynamic elements emanating from a common region are
strongly correlated if there is no real time connection. The scatter
seen in the correlation (see Figure 1) includes the variation of
$L_{bol}$ and $\alpha_{EUV}$ from their long term time average value
and the degree of scatter indicates that in general these variations
are modest.

\begin{figure*}
\includegraphics[width=170 mm, angle= 0]{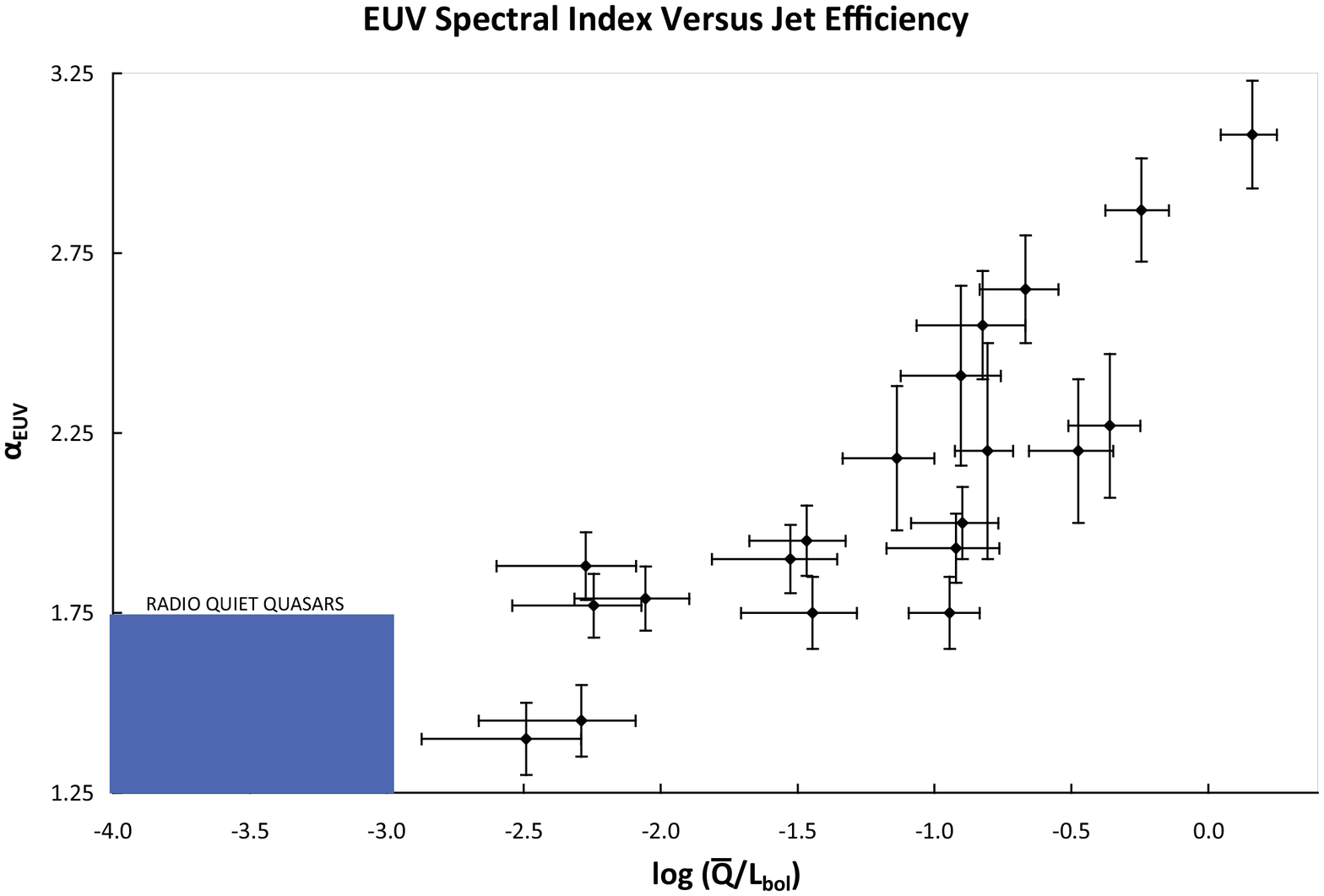}
\caption{The correlation of $\alpha_{EUV}$ versus the jet
efficiency, $\overline{Q}/L_{\mathrm{bol}}$ \citep{pun15,mar15}.
More powerful jets have depressed EUV emission.}
\end{figure*}

\par In this paper, a search is initiated for the real time
connection between $Q(t)/L_{bol}$ and $\alpha_{EUV}$. This is a
daunting task since it is not trivial to find an estimator that
scales with $Q(t)$ that is reliable and there is simply not enough
observing time on the \emph{Hubble Space Telescope} (HST) to monitor
quasar continua (spectra are needed to excise the emission lines) in
the EUV. Sufficient monitoring requires observing a source with
earth-based telescopes which requires $z> 3$ in order to get the EUV
redshifted into the U or B bands. This is not ideal due to the large
flux attenuation by the intervening Ly$\alpha$ forest. Firstly, the
quasar must have an incredibly large intrinsic EUV luminosity in
order to shine bright enough to be seen easily through the absorbing
screen. The second requirement is that the radio luminosity should
be high enough so that the object appears frequently in survey work
so that there is an ample database to cull through for
quasi-simultaneous observations. The third requirement is almost
mutually exclusive from the first two. In order to monitor $Q(t)$ by
radio flux, the source cannot suffer from Doppler aberration. This
is necessary since small variations in geometry would appear as
significant changes in the observed flux and this would provide a
false indicator of changes in the intrinsic jet luminosity
\citep{lin85}. Most of the radio sources at $z>3$ in the older radio
catalogs have sufficient flux density because they are Doppler
enhanced. Thus, we have a very strong set of restrictions due to the
high redshift, luminosity and lack of Doppler beaming: the source
needs an enormous intrinsic radio luminosity with no significant
Doppler beaming (i.e. $\sim$ 1Jy at $z>3$ with no Doppler beaming).
Do such sources exist? They seem to in the form of gigahertz peak
radio sources (GPS). Some of these at high redshift are incredibly
luminous \citep{ode98}. We argue below that the quasar 1442+101 at z
= 3.55 is a suitable candidate.

\begin{figure*}
\includegraphics[width=170 mm, angle= 0]{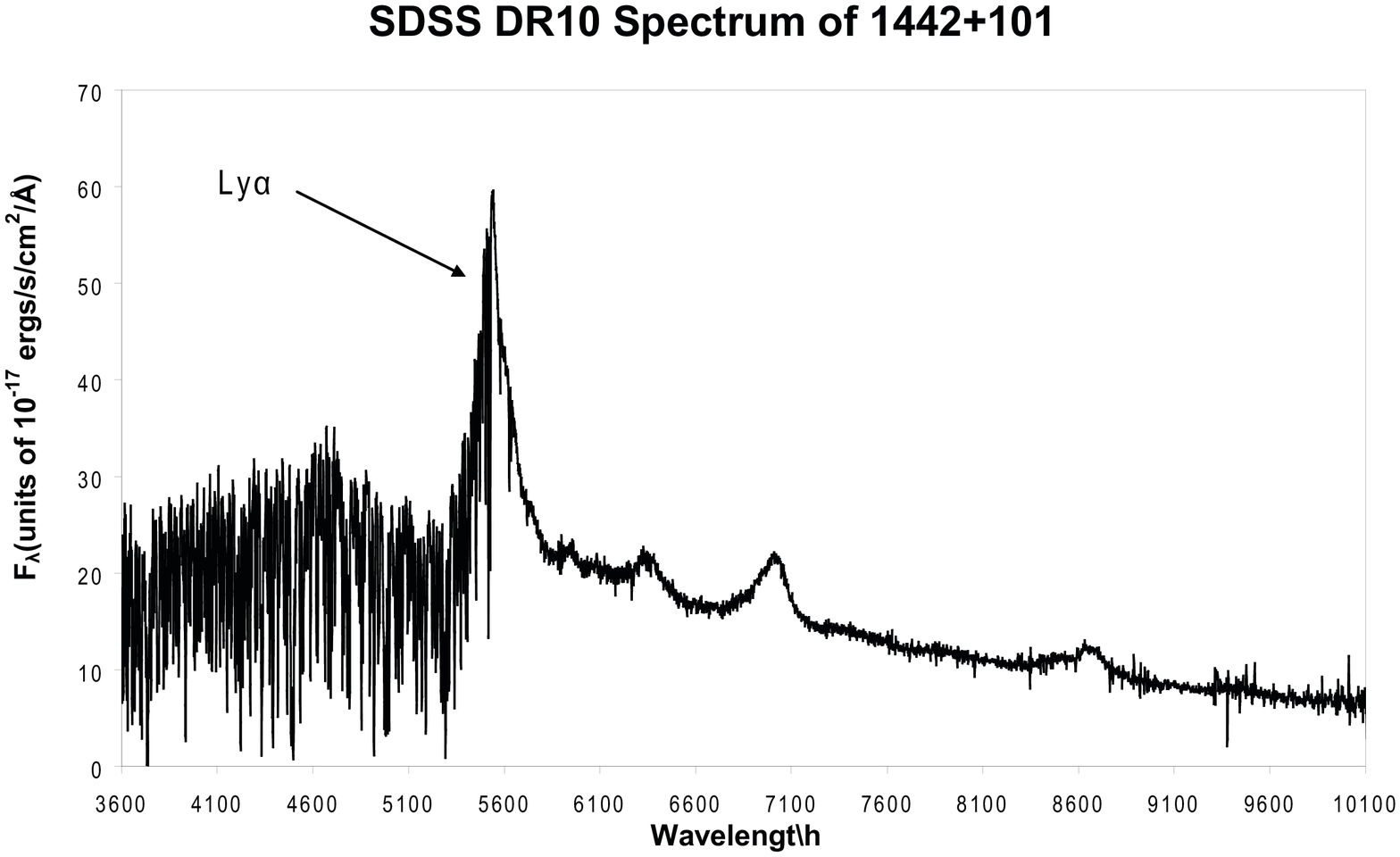}
\caption{Note the ample flux shortward of Ly$\alpha$ in the SDSS DR
10 spectrum of 1442+101.}
\end{figure*}

\section{Source Selection} The high redshift quasar, 1442+101, is a
powerful radio source that appeared in many of the early radio
surveys. The spectrum peaks at $\approx$ 1 GHz at 2.5 Jy. Therefore,
it was very surprising that the source was at extremely high
redshift. It was also surprising that unlike other high redshift
sources that appear in radio surveys, it is not blazar-like in that
it is not highly variable at radio frequencies.
\subsection{Doppler Beaming}It is argued here that the radio emission at
frequencies below 10 GHz is not Doppler beamed in 1442+101. A list
of evidence of Doppler beaming occurs below. For each item in the
list, the actual circumstance observed in 1442+101 is listed after
the colon.
\begin{enumerate}
\item Extreme radio variability: Not evident in the large dataset
compiled in the NASA Extragalactic Database or the monitoring in
\citet{tin03,min12}. This is indicative of non-blazar radio
emission.
\item Superluminal motion on 30 pc - 100 pc scales: There is no
measurable change in position of the three components seen at 1.6
GHz VLBI images in 1989 and in the 2.3 GHz VLBI images from 1999 on
scales of 10 pc - 100 pc \citep{dal95,pus12}. No blazar-like pc
scale evolution has been observed.
\item Flat spectrum radio core to $ > 30 $ GHz in the quasar rest
frame: The spectrum is very steep with a spectral index $\alpha> 1$
above 30 GHz in the quasar rest frame \citep{kov99,man09}. The
spectrum continues to steepen towards 1000 GHz in the quasar rest
frame \citep{ste95}. Thus, there is no ``buried" flat spectrum core
embedded within the source of the gigahertz peaked emission. There
is insignificant blazar-like synchrotron self-absorption at high
frequency.
\item Large optical variability: Low optical variability is reported
\citep{pic88,smi93}. The optical variability is similar to radio
quiet quasars and thus not typical of blazars.
\item Large optical polarization: No measurements exist.
Inconclusive evidence of blazar activity.
\end{enumerate}
The preponderance of evidence indicates that Doppler beaming does
not affect the emission from 1442+101 significantly.
\subsection{Estimator for $Q(t)$} If a source is not Doppler beamed
then it seems reasonable to choose the optically thin radio
luminosity created in the most compact regions of the source as a
quantity that scales with jet power. This emanates inside of any
working surface (a region of intense jet dissipation such as an
interaction with a dense cloud) and without synchrotron - self
absorption, we are looking directly at the synchrotron emission that
directly reflects dynamical elements, the energy of electrons and
the energy of the magnetic field in the jet. It might not be a
linear relationship, but increases in the optically thin radio
emission from the innermost jet should be positively correlated with
the jet power. The quasar 1442+101 is one of the most compact radio
sources known \citep{van84}. There is no emission, even in deep
observations, on kpc scales \citep{mur93}. Based on VLBI (very long
baseline interferometry) images of GPS quasars at lower redshift
(higher spatial resolution than is possible with 1442+101
observations), the radio emission is highly resolved with a small
fraction in the core unlike blazars \citep{sta97,sta01}. Most of the
emission especially at lower frequencies comes from knots in the jet
$>$ 10 pc from the core, the putative working surface. This is the
primary source of the optically thick low frequency emission in most
GPS quasars and likely 1442+101 as well. To estimate $Q(t)$, it is
not appropriate to sample the optically thick radio emission from
the ``working surface" that depends on dissipation that is
determined by dynamics and physical conditions external to the jet.
Above 8.4 GHz (observer frame) the spectral index is $\alpha> 1$
(optically thin) and this frequency is sampled frequently
\citep{kov99,man09,min12}. Thus, the 8.4 GHz flux density should be
an indicator of jet strength well inside the working surface. VLBI
imaging at 8.6 GHz indicates that the majority of the 8.4 GHz flux
is unresolved inside of 50 light years from the source
\citep{pus12}. This is the surrogate to monitor jet power given the
insufficient monitoring at higher frequencies.
\subsection{EUV Penetration of the Lyman Forest} The most
startling aspect of 1442+101 is that the EUV emission is so bright
that it shines through a dense forest of Ly$\alpha$ absorbing
clouds. The spectrum in Figure 2 indicates that there is more than
adequate luminosity to sample the spectrum at $\lambda_{o} = 4000$
\AA\, (observer frame) corresponding to $\lambda_{e} = 880$ \AA\, in
the quasar rest frame. The flux measured is not the intrinsically
emitted flux, but that attenuated by the Lyman forest. About five or
six absorbers have been resolved spectroscopically in a
$\delta\lambda_{o} = 100$ \AA\, window centered at $\lambda_{o} =
4000$ \AA\, \citep{bal74,bar90}. The intervening gas is comprised of
many small absorbing clouds, so one does not expect significant time
variation in the total flux integrated over a $100$ \AA\, window due
to changes in the total absorbing column. By measuring the EUV
continuum at $\lambda_{e}=880 \pm 11$ \AA\,we avoid contamination by
emission lines \citep{tel02}. The $100$ \AA\, ``smoothing" window
minimizes any affects caused by spectral resolution differences
between observations and the narrow absorption lines, allowing for a
consistent method of determining the observed flux density at
$\lambda_{o} = 4000$ \AA.

\begin{table}
 \centering
\caption{Quasi-simultaneous EUV and Radio Variability of 1442+101}
{\small
\begin{tabular}{ccccc}
 \hline

 Observation & Flux Density      & Flux Density      & Telescope & Delta\tablenotemark{a}     \\
    Date (MJD)     &  8.46 GHz (mJy)    &  $ \lambda_{o} = 4000$ \AA ($\lambda_{e} = 880$ \AA)      & Radio/ Optical         &     \\
 Radio/ Optical    &                   &  ($10^{-17}$         &                     &  (Days)   \\
                   &                   &   ($\mathrm{ergs/sec/cm^{2}/\AA})$   &                     &    \\
\hline
42126/ 42189             &  $ 598 \pm 42 $ \tablenotemark{b}   & $ 23.4 \pm 2.4$\tablenotemark{c} & Parkes 64m/ Lick 3m  & 297    \\
47266/ 47320              &  $ 709 \pm 18 $\tablenotemark{d}    & $ 15.3\pm 2.3 $\tablenotemark{e}  & VLA/  Hale 5m        & 54    \\
48415/ 48427              & $ 704 \pm 18 $\tablenotemark{d}     &  $ 16.4\pm 1.2 $ \tablenotemark{f} &    VLA/ Lick 3m      & 12    \\
53940/ 53827              &  $ 663 \pm 16 $ \tablenotemark{d}   &  $ 21.3\pm 1.3 $\tablenotemark{d}  &   VLA/ SDSS DR7       & 113   \\
55834/ 55976              &  $ 659 \pm 16 $\tablenotemark{d}    &  $ 20.5\pm 1.4 $\tablenotemark{d} &   VLA/ SDSS DR10       & 142   \\
\hline
\end{tabular}}
\tablenotetext{a}{Number of days between radio and optical
observations} \tablenotetext{b}{\citet{shi81}}
\tablenotetext{c}{\citet{bal74}} \tablenotetext{d}{This paper}
\tablenotetext{e}{\citet{bar90}} \tablenotetext{f}{\citet{lyo95}}
\end{table}

\section{Results} There are far more radio observations than
calibrated optical spectra down to $\lambda_{o} = 4000 \AA$. The
optical and radio variability is small \citep{pic88, tin03}. Due to
the paucity of optical data sampling it thus seems reasonable (but
not ideal) to consider observations within 300 days as
quasi-simultaneous. Due to cosmological redshifting, this
corresponds to about 66 days in the quasar rest frame. To put this
in perspective, we estimate the size of the putative active
EUV/radio emitting region in 1442+101 as follows. Based on standard
bolometric corrections, Figure 2 and archival IR (rest frame
optical) data from the NASA Extragalactic Database, we estimate
$L_{bol} \approx 4 - 5 \times 10^{47}$ ergs/sec, ignoring the
contribution from reprocessed emission (avoiding double counting) in
the molecular clouds \citep{pun15,dav11}. If the accretion flow is
radiating at $<50\%$ of the Eddington rate, the black hole mass
would be $M_{bh}
> 8\times 10^{9} M_{\odot}$. For a
rapidly spinning black hole, 66 days corresponds to $< 2$ Keplerian
orbital periods at a distance of $5 R_{g}$ ($R_{g} > 1.1\times
10^{15}$ cm is the black hole radius for a rapidly rotating black
hole) from the central black hole in the equatorial plane. This also
corresponds to $< 4$ Keplerian orbital periods at a distance of $3
R_{g}$ from the central black hole in the equatorial plane. These
are reasonable estimates for the location of the EUV emitting region
from the variability of the EUV and numerical models of accretion
flows \citep{mar97,hab03,zhu12,pun14,pun15}. Thus, 300 days is not
that unreasonable of a condition for quasi-simultaneity in the
context of jet production and EUV emission.
\begin{figure*}
\begin{center}
\includegraphics[width=120 mm, angle= 0]{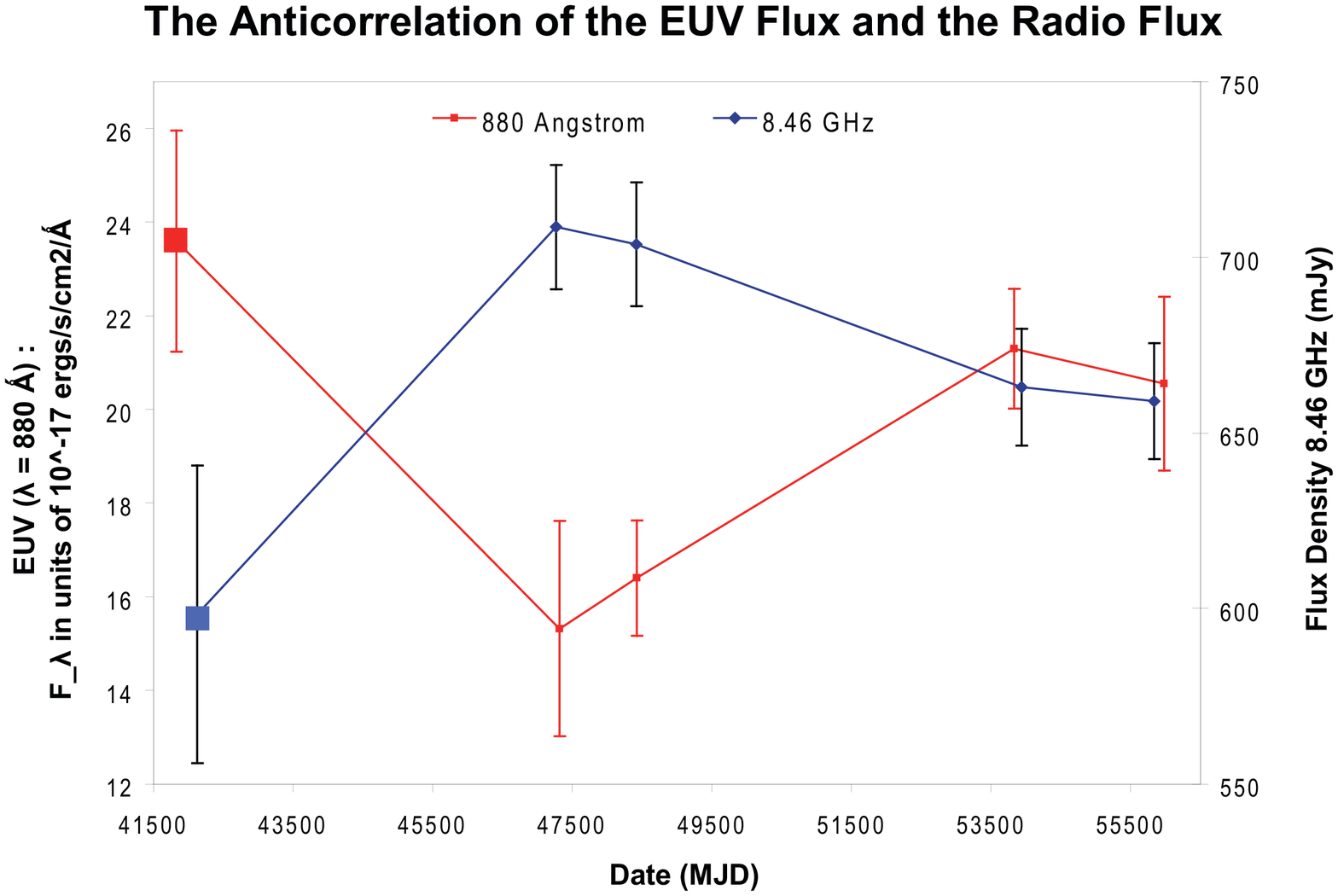}
\includegraphics[width=120 mm, angle= 0]{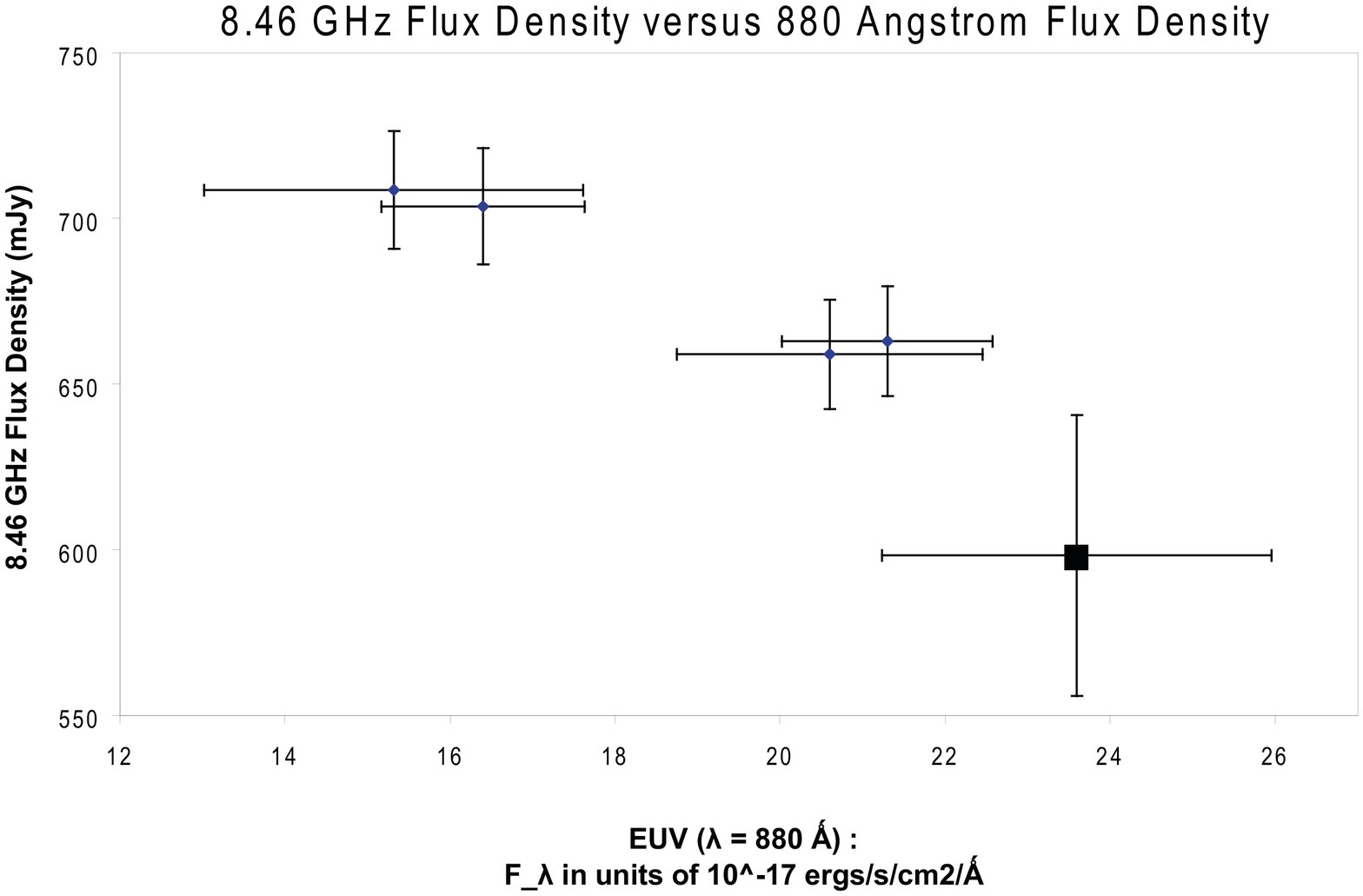}
\caption{The top frame shows the light curves for EUV flux density
and 8.46 GHz flux density sampled at quasi-simultaneous observation
epochs. The bottom frame shows the anti-correlation of the EUV flux
density and the 8.46 GHz flux density. The squares indicate the
observational epoch that is not in the ``gold sample" (not a VLA
radio observation) described in the text.}
\end{center}
\end{figure*}
\begin{figure*}
\includegraphics[width=120 mm, angle= 0]{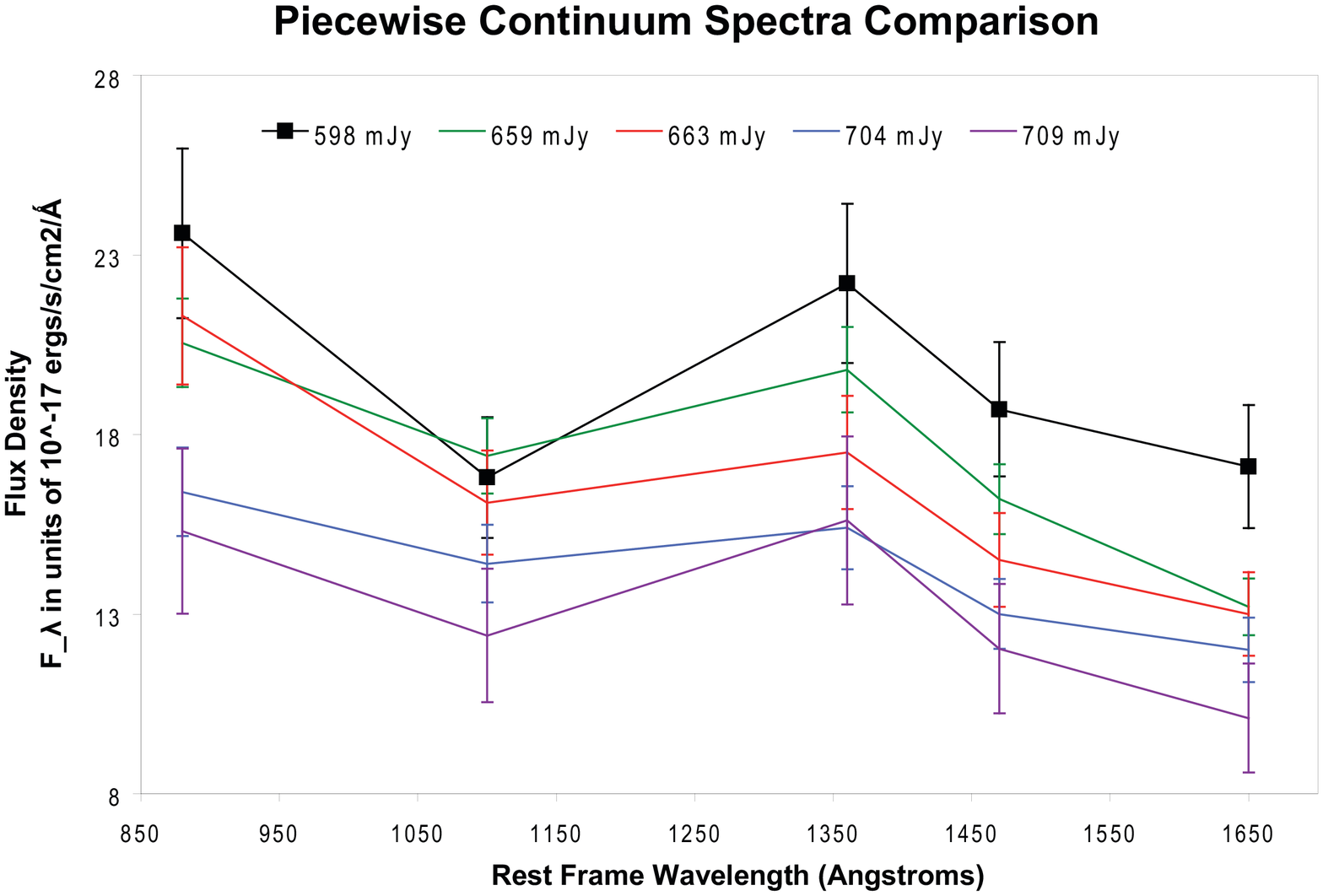}
\caption{The piecewise continuum estimates in the far UV and EUV of
the different epochs labeled by their 8.46 GHz flux density. The
squares indicate the observational epoch that is not in the ``gold
sample" (not a VLA radio observation) described in the text. The
continuum excess of weak jet states relative to strong jet states is
largest in the EUV.}
\end{figure*}

\par Table 1 shows the data for the 5 quasi-simultaneous epochs that
are spread out over $\sim 40$ years (1973 - 2012). Table 1 lists the
date of observations (column 1), the 8.46 GHz flux density (column
2), the flux density at $\lambda_{o} = 4000 \AA$ (column 3), the
telescopes (column 4), the difference in the observation date from
the optical and radio in days (column 5) and the references (columns
6 and 7). For observations that are at a slightly different
frequency than 8.46 GHz ($\sim$ 100 - 400 MHz), a spectral index of
1 is assumed based on the MJD 47266 VLA observations at 8.21 GHz and
8.66 GHz and archival data \citep{kov99,min12,man09}. We note that
the observation in \citet{bar90} on MJD 47320 in Table 1 is chosen
to have the largest fractional uncertainty, 15\%, of all the epochs.
Issues with the flux calibration have been noted based on the
atmospheric dispersion correction for the \citet{bar90} sample
\citep{cor91}. For SDSS DR10 quasar data, such as the MJD 55976
observation in Table 1, there is a calibration issue in the blue
part of the spectrum. Specifically, in order to increase spectral
sensitivity for quasars, the hole drilled over the aperture is
centered on the displacement associated with 4000\AA. However, the
calibration of the star has a hole centered about the displacement
associated with 5400\AA. This inconsistency artificially raises the
flux level in the blue that we estimate as a 7.2\% excess for quasar
light at 4000\AA\, (from the airmass of 1.1 and seeing  of 1.4
arcsec). If there is a similar underestimate of the standard star
flux at 4000\AA\,, the total estimated increase in the DR10 quasar
spectral flux is 14.4\% at 4000\AA. As a check of this calculation,
we note that this agrees with the offset seen between the
photometric g magnitude and the synthetic magnitude computed from
the  spectrum (18.61 and 18.39, respectively). The corrected
spectrum is plotted in Figure 2.

\par Table 1 can be used to assess the real time
connection between $Q(t)/L_{bol}$ and $\alpha_{EUV}$. Notice that,
per the discussions by \citet{pun15}, $L_{bol} \approx 3.8
F_{\lambda_{e}}(\lambda_{e} = 1100 \AA)$ in the quasar rest frame
and $ \alpha_{EUV} = \log{\frac{F_{\lambda_{e}}(\lambda_{e} = 880
\AA)}{F_{\lambda_{e}}(\lambda_{e} = 1100
\AA)}}/\log{\frac{1100\AA}{880\AA}}$. \footnote{Our spectra do not
extend all the way to $\lambda_{e} = 700$ \AA\,. A power law fits
the continuum well in this range for other quasars \citep{pun15}.
Using $\lambda_{e} = 880$ \AA\, to compute the power law should be
an adequate expedience to see changes in spectral slope.} Thus,
$\alpha_{EUV}$ varies monotonically with
$F_{\lambda_{e}}(\lambda_{e} = 880 \AA)/F_{\lambda_{e}}(\lambda_{e}
= 1100 \AA)$. If there is a relationship between $Q(t)/L_{bol}$ and
$\alpha_{EUV}$ then there is a relationship between $Q(t)/L_{bol}$
and $F_{\lambda_{e}}(\lambda_{e} = 880
\AA)/F_{\lambda_{e}}(\lambda_{e} = 1100 \AA)$. The advantage of the
latter relationship is that both quantities have
$F_{\lambda_{e}}(\lambda_{e} = 1100\AA)$ in the denominator. Thus if
one multiplies both quantities by $F_{\lambda_{e}}(\lambda_{e} =
1100 \AA)$, one only needs to look at the light curves of $Q(t)$ (as
traced by the 8.46 GHz flux density) and
$F_{\lambda_{o}}(\lambda_{o} = 4000 \AA)$ (equivalently
$F_{\lambda_{e}}(\lambda_{e} = 880 \AA)$) in order to detect a
correlation between $Q(t)/L_{bol}$ and $\alpha_{EUV}$. The
motivation for this re-normalization is to remove the undesirable
statistical scatter caused by dividing both the quantities by
$F_{\lambda_{o}}(\lambda_{o} = 5000 \AA)$ (corresponding to
$\lambda_{e} = 1100$\AA\,) that adds to each measured quantity
considerable uncertainty, but does not add physical content. Such an
expedience would not be justified in an analysis that comprised
multiple objects since each object is normalized differently and the
normalization is time variable.
\par The top frame of Figure 3 is a plot of both 8.46 GHz flux density
and $F_{\lambda_{o}}(\lambda_{o} = 4000 \AA)$ as a function of time.
The bottom frame of Figure 3 shows that $Q(t)$ decreases as
$F_{\lambda_{o}}(\lambda_{o} = 4000 \AA)$ increases as expected.
Even though the number of observations is too small for rigorous
statistical analysis, we note that Pearson correlation coefficient
is -0.949 corresponding to a one-sided 99.3\% statistical
significance. If one restricts the analysis to a ``gold sample" of
epochs all observed with the same telescope (the VLA) and within 150
days of the optical observation ($<2$ Keplerian orbits $3R_{g}$ from
the black hole), one still sees a correlation. The two stronger
radio states have the weakest $F_{\lambda_{o}}(\lambda_{o} = 4000
\AA)$ and vice versa for the two weaker radio states.
\par In the analysis of HST composite spectra in
\citet{pun14,tel02}, a deficit of EUV continuum emission was
associated with the radio loudness in quasars. There was no deficit
of emission seen in the far UV continuum for RLQs. However, the
continua plotted in Figure 4 indicate a far UV continuum suppression
as the jet gets stronger. The plot shows piecewise continuum spectra
estimates in the far UV and EUV at the different epochs labeled by
their 8.46 GHz flux density. The rest frame points were sampled at
wavelengths in which line emission seemed minimal in all the
spectra. Rest frame wavelengths were indicated for ease of
comparison to well known emission lines. The observed wavelengths
are 4000 \AA\,, 5000 \AA\,, 6200 \AA\,, 6700 \AA\,, and 7500 \AA\,.
There are three radio ``weaker" jet states at 598 mJy, 659 mJy, 663
mJy and two ``stronger" jet states at 704 mJy and 709 mJy. The UV
fluxes are highly clustered at $\lambda_{e} = 1650$\AA\,, four of
the flux density measurements are not statistically significantly
different. The flux density of the continua associated with the weak
jet states exceeds that of the continua associated with the strong
jet states by an ever increasing differential as the the wavelength
decreases. Furthermore, the EUV spectra from 1100\AA\, to 880\AA\,
are softer (larger $\alpha_{EUV}$) for the stronger jet states than
for the weaker jet states as expected from a $Q(t)/L_{bol}$ and
$\alpha_{EUV}$ correlation. The continuum suppression associated
with jet power might taper off in the far UV as opposed to an abrupt
turn off at $\lambda_{e} = 1100$\AA\,.
\section{Discussion} The physical interpretation of the correlation
between jet power and the decrement in EUV luminosity that is
illustrated in Figure 1, 3 and 4 was elucidated in two previous
studies \citep{pun14,pun15}. In this section, we review this
analysis and discussion in order to put the results into
perspective. The most straightforward explanation of the correlation
involves jets from magnetic flux in the inner accretion disk where
the EUV emission also originates. As more poloidal vertical magnetic
flux is stored in the innermost accretion disk, the volume available
for the thermal emitting gas is displaced. More poloidal magnetic
flux equates to a stronger jet and less optically thick thermal gas
equates to a weaker EUV luminosity. This dynamical configuration is
known as magnetically arrested accretion \citep{igu08}.

\par Many different dynamical systems have been classified as magnetically arrested in the
numerical simulation literature. The relevant numerical simulations
for RLQs are the ones in which the vertical magnetic flux is
distributed in magnetic islands of low plasma density that are
concentrated in the inner accretion flow. The configuration is far
from time stationary. The islands are dynamic and buoyantly move
through the ram pressure imposed by the dense thermal EUV emitting
plasma by means of a series of Kruskal-Schwarzschild instabilities
\citep{igu08}. The ram pressure of the accretion flow concentrates
magnetic flux near the black hole.

\subsection{A Review of Evidence for Magnetically Arrested
Accretion in RLQs} Evidence of the detailed predictions of
magnetically arrested accretion was found in \citet{pun15} by
exploring a basic model. The first prediction is that due to the ram
pressure interaction of the accreting gas with the magnetic islands,
the correlation of $\overline{Q}/L_{bol}$ and $\alpha_{EUV}$ should
be stronger than the correlation between $\overline{Q}$ and
$\alpha_{EUV}$. This was verified by a partial correlation analysis
that indicated that the correlation of
$\overline{Q}/L_{\mathrm{bol}}$ with $\alpha_{\mathrm{EUV}}$ is
statistically significant and the correlation of $\overline{Q}$ with
$\alpha_{\mathrm{EUV}}$ is spurious.

\par The second piece of evidence supporting magnetically arrested accretion
that was found in \citet{pun15} is a verification of the specific
relationship between the jet power and the EUV suppression that
arises from the basic model of magnetically arrested accretion. The
derivation will not be repeated here, but the introduction of some
notation is required to explain the exact result that was
demonstrated. If the model is representative of the physical
circumstance depicted in Figure 1 then the filling fraction of the
inner accretion disk with magnetic islands, $f$, is related to the
deficit of EUV luminosity, where the EUV luminosity deficit is
evaluated relative to luminosity of the EUV continuum of radio quiet
quasars (RQQs). In order to compare the EUV deficit from object to
object, the EUV spectral luminosity was normalized in \citet{pun15}
to the spectral luminosity at the approximate peak of the spectral
energy distribution (SED),
\begin{equation}
\mathrm{Normalized \; EUV\; Spectral \; Luminosity} \equiv
\frac{L_{\nu}(\lambda= 700\AA)}{L_{\nu}(\lambda= 1100\AA)}\;,
\end{equation}
where $L_{\nu}(\lambda= 700\AA)$ is the EUV continuum spectral
luminosity and $L_{\nu}(\lambda= 1100\AA)$ is the spectral
luminosity at the approximate peak of the SED. Defining the EUV
deficit of a RLQ relative to RQQs requires defining a fiducial
normalized EUV spectral luminosity for RQQs from composite spectra
\citep{pun15}
\begin{equation}
\mathrm{Fiducial\; Normalized \; EUV\; Spectral \; Luminosity}
\equiv \frac{L_{\nu}(\lambda= 700\AA)}{L_{\nu}(\lambda=
1100\AA)}\biggr\vert_{\mathrm{RQQ}} \;,
\end{equation}
where $L_{\nu}(\lambda= 700\AA)$ is the EUV continuum spectral
luminosity of the RQQ composite spectrum and $L_{\nu}(\lambda=
1100\AA)$ is the spectral luminosity at the approximate peak of the
SED in the RQQ composite spectrum.
 The normalized EUV deficit is
approximately equal to the magnetic flux fill fraction, $f$, of the
inner disk (equivalently the fractional volume of EUV emitting gas
that is displaced by magnetic flux tubes) in each individual RLQ
\begin{equation}
\mathrm{EUV \, Deficit}\equiv \left[1 - \left[\frac{L_{\nu}(\lambda=
700\AA)}{L_{\nu}(\lambda=
1100\AA)}\right]\left[\frac{L_{\nu}(\lambda=
1100\AA)}{L_{\nu}(\lambda=
700\AA)}\right]\biggr\vert_{\mathrm{RQQ}}\right] \approx f \;.
\end{equation}
Since the jet power scales with the square of the poloidal magnetic
flux contained within the jet base, the scaling that is predicted by
the basic model of magnetically arrested accretion was shown by
\citet{pun15} to be of the form
\begin{equation}
\overline{Q}/L_{\mathrm{bol}} \approx Af^{2}\approx A\left[1 -
\left[\frac{L_{\nu}(\lambda= 700\AA)}{L_{\nu}(\lambda=
1100\AA)}\right]\left[\frac{L_{\nu}(\lambda=
1100\AA)}{L_{\nu}(\lambda=
700\AA)}\right]\biggr\vert_{\mathrm{RQQ}}\right]^{2} \;,
\end{equation}
where A is a constant. The best fit to the RLQ data was found to be
\begin{equation}
\overline{Q}/L_{\mathrm{bol}} = A\left[1 -
\left[\frac{L_{\nu}(\lambda= 700\AA)}{L_{\nu}(\lambda=
1100\AA)}\right]\left[\frac{L_{\nu}(\lambda=
1100\AA)}{L_{\nu}(\lambda=
700\AA)}\right]\biggr\vert_{\mathrm{RQQ}}\right]^{2.05 \pm 0.17} \;.
\end{equation}
The uncertainty in the exponent includes uncertainty in the fiducial
RQQ normalized EUV spectral luminosity in Equation (2) as well as
the statistical scatter in the data. The agreement of the exponent
in equations (4) and (5) within these uncertainties provides strong
evidence that the mechanism that is creating the jet and suppressing
the EUV emission is in fact similar to magnetically arrested
accretion in the innermost accretion disk. It is further estimated
from the same analysis that $f$ is a few percent for the RLQs with
weaker jets and $\sim 50\%$ for the RLQs with the most powerful jets
\citep{pun15,mar15}.
\subsection{Review of the EUV Deficit in the Context of Numerical Simulations}
A major shortcoming of 3-D numerical models of accretion onto black
holes is that the magnetic field topology and poloidal flux
distribution is determined by physics beyond the assumed single
fluid ideal magnetohydrodynamic (MHD) approximation. This is
critical because it has been shown that even small changes in the
numerics near the black hole will drastically alter the poloidal
magnetic flux topology and the entire dynamical scheme
\citep{pun16}. Most prominent amongst these missing features is the
microphysics that determines the diffusion rate of plasma onto and
off of magnetic field lines and the magnetic field reconnection
rate. Neither of these are well known in the exotic environment near
black holes. These physical processes are critical not only for the
formation of the magnetic islands, but the time evolution of the
magnetic islands is determined primarily by diffusion
\citep{igu08,pun16}. Even worse, these dynamical elements occur only
as a consequence of numerical diffusion in modern simulations in
over simplified ideal MHD single fluid models of the physics
\citep{pun16}. Future development is difficult since it is not even
clear theoretically what the basic principles required for an
accurate physical depiction would be. Many issues that are related
to these topics are active areas of investigation in solar and
fusion physics \citep{bau13,mal09,thr12,yam07}. As such it is
imperative that observational results such as those presented here
are needed to guide the course of numerical and theoretical work.
\par The observational evidence provided by this study can be used to cull through possible
numerical and physical scenarios in order to see which ones are
viable candidates to represent the physical system in a RLQ. The
following restrictions for numerical work are model independent and
derive directly from the observations \citep{pun15}. These can be
used to guide us toward numerical models that replicate the physics
of jet launching in RLQs. Empirically, it is determined that both
the large scale poloidal magnetic flux at the base of the jet and a
mechanism that suppresses (but does not eliminate) the EUV emission
from the innermost accretion flow coexist at the heart of the
central engine of RLQs. This is too large of a coincidence, some of
this magnetic flux must be the same element that disrupts, but does
not eliminate the innermost accretion flow. This implies that there
is significant vertical poloidal magnetic flux in the inner
accretion flow of RLQs. If a numerical effort cannot reproduce this
circumstance then perhaps the numerical approximation to the
relevant physical processes requires further development.
Observations are now mature enough to lead the numerical work.
\par Summarizing the status of the numerical work, it is noted that the
magnetically arrested accretion discussed above in the introduction
to this section is qualitatively similar to the 3-D numerical
simulations explored in \citet{igu08} and \citet{pun09}. They
produce a fill fraction, $f$, in the innermost accretion flow that
is consistent with the range described below Equation (5). They also
provide most of the gross features of time evolution required to
support the jets and suppress the EUV emission in the innermost
accretion flow \citep{pun15}.  By contrast, the simulations in
\citet{mck12,tch11,tch12} that are heavily seeded with large scale
magnetic flux are devoid of magnetic islands close to the event
horizon. This is evidenced by the claim in \citet{mck12} that no
significant Poynting flux emerges from this region as well as the
linked online videos of the simulations. The videos show the
innermost significant, modest, magnetic island concentrations are
located at $r > 10R_{g}$ and they are extremely transient. In
\citet{mck12,tch11,tch12} the fill fraction, $f\approx 0$ in the
innermost accretion flow. This is basically a zone of avoidance for
vertical poloidal magnetic flux in these simulations. Instead of
penetrating the thickness of the equatorial accretion flow,
vertically, the poloidal field spreads out above and below the
accretion disk in radial fans. Thus, these simulations are not
consistent with the observations of RLQs. These radial fans compress
or ``choke" the flow. The ``magnetically choked accretion flows" of
\citet{mck12} heat the innermost accretion flow compressively as
part of the magnetic choking process and likely increase radiation
from this region. This is verified by the recent results from this
family of simulations \citep{ava15}. The inner accretion flow
luminosity is claimed to increase significantly over a standard
optically thick, geometrically thin accretion disk, the opposite of
what is observed in RLQs.
\subsection{The EUV Deficit and Jet Formation Across all Scales}
\par It is tempting to put these results into a broader context.
It is possible that all black hole related jet phenomena are a
consequence of vertical magnetic flux impeding the inner accretion
flow. In particular, for Galactic black hole binaries (GBH) jet type
emission occurs only when the soft X-ray emission (the putative
thermal emission from the accretion disk) is suppressed
\citep{rus11,fen04,kle02}. However, the ``coronal" X-ray power law
emission is not suppressed and can be very strong
\citep{pun13,fuc03}.

\par There have also been claims that dips in the coronal X-ray flux
(the X-ray power law) are coordinated with superluminal ejections in
the Seyfert galaxy, 3C120 \citep{mar02,cha09,loh13}. If one were to
make an analogy between the GBHs and 3C120, the superluminal
ejections in 3C 120 would have to be the equivalent of the discrete
superluminal ejections in GBHs. However, superluminal discrete
ejections in GBHs are characterized by very strong coronal emission
in the last hours just before ejection that reach near historic
maxima that can exceed 50\% of the Eddington luminosity
\citep{pun13}. During the ejection, the X-ray luminosity becomes
highly variable with an average value similar to that just before
the ejection \citep{pun13}. Thus, the claim of coronal X-ray power
law luminosity dips in 3C 120 during superluminal ejections refers
to a different phenomenon.

The corona X-ray power law emission during superluminal ejections in
3C 120 likely emerges from close to the black hole. However, there
is no direct evidence that the emission from the innermost optically
thick accretion flow is suppressed during superluminal ejections.
The EUV emission from the high frequency tail of the optically thick
thermal emission is not monitored. As noted above, in GBHs, the
X-ray power law luminosity is not correlated with the optically
thick thermal luminosity. Thus, there is no precedent that justifies
the assumption that a dip in the X-ray power law equates to a dip in
the optically thick thermal emission from the inner accretion disk
in 3C 120.  Thus, the physical state of the inner optically thick
accretion disk during superluminal ejections is unknown. There is a
claim of indirect evidence of inner disk movement outward during
superluminal ejections based on Fe K$\alpha$ line width arguments
derived from three X-ray observations \citep{loh13}. However, the
interpretation of the X-ray data in the ``high state" observed with
XMM and the inner disk location claimed by \citet{loh13} disagrees
with the seminal studies of that data \citep{bal04,ogl05}. The
original studies of the XMM ``high state" indicated that the Fe
K$\alpha$ line was produced in cold plasma at $ > 75 R_{g}$ from the
central black hole. This is very far from the innermost stable
circular orbit where the cold plasma should reside if there were
evidence of ``a complete disk extending down to the innermost stable
circular orbit (ISCO) during the XMM-Newton observation" as claimed
by \citet{loh13}.
\par The claim of evidence of an interaction of the jet with the corona during superluminal ejections in 3C 120 is not well
understood, primarily because the corona is not a well understood
region. There is strong evidence that the corona might be an
out-flowing wind in Seyfert galaxies \citep{liu14,kha15}. Does the
corona envelope the disk like a stellar corona, or is it a separate
optically thin radiatively inefficient disk inside the accretion
disk, the so-called truncated disk scenario (e.g., \citet{loh13})?
The physical interpretation of the depressed coronal emission is
highly dependent on the model of the corona.

In summary, both GBHs and RLQs show a suppression of the optically
thick thermal emission when jetted phenomenon occur. The Seyfert
galaxy, 3C 120, shows something similar, but it is the coronal
emission that appears suppressed which does not occur in GBHs. It
would be a major advance if these three phenomena could be
connected. There are many poorly understood details that would need
to be studied and clarified further before we are at that point.

\section{Conclusion and Future Prospects} This paper studies
archival data on the distant quasar, 1442+101. Evidence of a real
time connection between the jet power and the suppression of EUV
emission is found. The result is based on only 5 epochs and
simultaneity was established only within a 300 day interval. These
findings motivate the need for a long term monitoring program of
this source and other high redshift GPS. The radio observations
would be improved with multi-frequency observations with VLA or
ATCA. Knowing the flux density at more than one frequency (perhaps
8.4 GHz, 15 GHz and 22 GHz) would allow one to compute the
luminosity of the optically thin emission and would be less prone to
errors induced from slight changes in spectral slope that are
problematic with a single frequency proxy. Coordinated sampling two
times a year would be sufficient. Ideally, the source should be
monitored with the same optical and radio instrument to minimize
uncertainties in this difficult measurement. The VLA ``gold sample"
implemented here improves the situation, but there are still
multiple optical telescopes involved.
\begin{acknowledgements}
We would like to thank Tracy Clarke for supplying the VLA data from
September 30, 2011. We would also like to thank Patrick Ogle and
Robert Antonucci for their valuable analysis of the 3C 120 X-ray
data. BP notes that this research was supported by ICRANet.
\end{acknowledgements}

\end{document}